\documentclass[letterpaper]{article} 
\usepackage{aaai24}
\usepackage{times}  
\usepackage{helvet}  
\usepackage{courier}  
\usepackage[hyphens]{url}  
\usepackage{graphicx} 
\usepackage{natbib}  
\usepackage{caption} 
\usepackage{svg}
\usepackage{soul}
\usepackage[utf8]{inputenc}
\usepackage[T1]{fontenc}
\usepackage{amsthm}
\usepackage{amsmath}
\usepackage{amsfonts} 
\usepackage{subfigure}
\usepackage{booktabs}
\usepackage[ruled,commentsnumbered,linesnumbered]{algorithm2e}
\usepackage{xspace} 
\usepackage{enumitem} 
\usepackage{tabularx} 
\usepackage{multirow}
\title{Meta-Stock: Task-Difficulty-Adaptive Meta-learning for Sub-new Stock Price Prediction}
\author{
    Linghao Wang\textsuperscript{\rm 1},
    Zhen Liu\textsuperscript{\rm 1},
    Peitian Ma\textsuperscript{\rm 1},
    Qianli Ma\textsuperscript{\rm 1,2}\thanks{Qianli Ma is the corresponding author.}
}
\affiliations{
    School of Computer Science and Engineering, South China University of Technology, Guangzhou China\textsuperscript{\rm 1}\\
    Key Laboratory of Big Data and Intelligent Robot (South China University of Technology), Ministry of Education\textsuperscript{\rm 2}\\
    wlhsama@gmail.com,
    cszhenliu@mail.scut.edu.cn,
    ma\_scuter@163.com,
    qianlima@scut.edu.cn,
}

\begin{document}

\maketitle

\begin{abstract}
    Sub-new stock price prediction, forecasting the price trends of stocks listed less than one year, is crucial for effective quantitative trading. While deep learning methods have demonstrated effectiveness in predicting old stock prices, they require large training datasets unavailable for sub-new stocks. In this paper, we propose Meta-Stock: a task-difficulty-adaptive meta-learning approach for sub-new stock price prediction. Leveraging prediction tasks formulated by old stocks, our meta-learning method aims to acquire the fast generalization ability that can be further adapted to sub-new stock price prediction tasks, thereby solving the data scarcity of sub-new stocks. 
    Moreover, we enhance the meta-learning process by incorporating an adaptive learning strategy sensitive to varying task difficulties. Through wavelet transform, we extract high-frequency coefficients to manifest stock price volatility. This allows the meta-learning model to assign gradient weights based on volatility-quantified task difficulty.
    Extensive experiments on datasets collected from three stock markets spanning twenty-two years prove that our Meta-Stock significantly outperforms previous methods and manifests strong applicability in real-world stock trading. Besides, we evaluate the reasonability of the task difficulty quantification and the effectiveness of the adaptive learning strategy.
\end{abstract}

\section{Introduction}
Sub-new stocks are stocks listed for less than one year. Compared to stocks listed for longer periods, the price trends of sub-new stocks are more volatile, allowing investors to profit from short-term trading \cite{subnew}. Consequently, predicting the price of sub-new stocks can be valuable for both stock traders and quantitative finance researchers. 

Due to short listing time, data scarcity is the main challenge for sub-new stock price prediction, and introducing supplement information is the most direct approach to address this issue. Previously, textual data such as social media information  \cite{NLP1,NLP2} and company relations extracted via graph neural networks \cite{GraphNLP} have been examined to facilitate stock price prediction. Although these methods can tackle limited data availability in theory, high-quality supplementary data remains difficult to obtain and assess \cite{dong2020belt,batra2018integrating}.

\begin{figure}
    \centering
    \includegraphics[width=0.9\columnwidth]{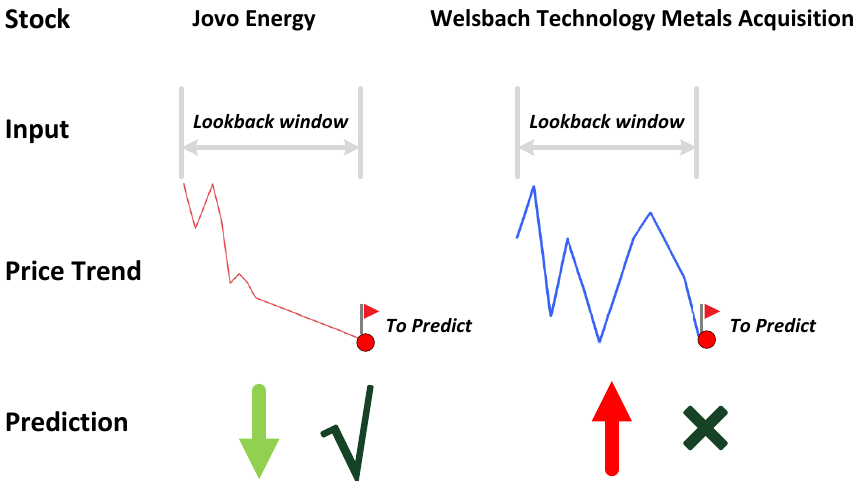}
    \caption{The price series of Welsbach Technology Metals Acquision reflect larger temporal volatility, and is harder to predict than the price trend of Jovo Energy. The red flag represents the date on which the model predicts the stock price trend.}
    \label{fig:intro}
\end{figure}

Apart from introducing supplement data, transfer learning and meta-learning are two machine learning techniques employed to tackle data scarcity. Specifically, transfer learning obtains a pre-trained model via large amounts of source data and fine-tunes the model on limited target domain data \cite{NeuralNetwork3}. Acquiring useful features via source data, transfer learning, to some extent, allows the fine-tuned model to solve target tasks with smaller datasets. However, when data distributions of source tasks and target tasks significantly differ, transfer learning underperforms since the pre-trained models may overfit source data and thus fail to adapt to target data \cite{yang2020bi}. In contrast to data-focused transfer learning, meta-learning is task-focused, emphasizing generalizing practical learning strategies instead of transferring low-level features \cite{meta-learning}. By learning "how to learn" on source tasks, meta-learning enables a model to quickly adapt and generalize to unseen tasks, relying less on task similarity and domain match. However, these two learning techniques have not been applied to sub-new stock price prediction, with their performance remaining unknown. Given the non-stationarity of stock price series \cite{zhang2017stock,wen2019stock}, significant differences exist between old and sub-new stock data. Therefore, meta-learning is a more reasonable approach for sub-new stock price prediction. 

However, applying meta-learning to sub-new stock price prediction still faces challenges. Firstly, meta-learning is a task-based approach, but most existing methods for stock price prediction are data-focused \cite{ACL22,Graph5,GraphNLP}, with little research on task construction for stock price prediction. Therefore, constructing tasks in stock price prediction contexts is crucial before employing meta-learning. Additionally, the difficulty levels of prediction tasks are disparate due to the varying volatility of price series. For example, as shown in Figure \ref{fig:intro}, the price trend of Welsbach Technology Metals Acquisition (higher volatility) may be more difficult to predict than that of Jovo Energy (lower volatility). However, traditional meta-learning performs equivalent training on each task \cite{MAML,meta-learning}, failing to deal with tasks based on their difficulty levels. This weakness undermines the effectiveness of meta-learning in capturing task-specific knowledge and acquiring fast generalization ability vital for sub-new stock price prediction. %

To address the above issues, we propose a task-difficulty-adaptive meta-learning model: Meta-Stock.
With numerous old stock price prediction tasks, Meta-Stock employs meta-learning to adapt the generalization ability acquired from these tasks to the sub-new stocks, overcoming sub-new stock data scarcity.
Besides this traditional meta-learning process, we incorporate an adaptive learning strategy to tackle disparate task difficulty levels, thereby enhancing meta-learning effectiveness. Specifically, task difficulty levels can be measured by price volatility in stock price prediction contexts \cite{xiang2022temporal}. Based on this assumption, we employ wavelet transform to measure the volatility of stock price series.
Instead of using wavelet transform to extract low-frequency components for capturing general price trends \cite{teng2020enhancing,luo2021study,wu2021hybrid}, we employ it to extract high-frequency coefficients manifesting irregular volatility \cite{lahmiri2014wavelet} and utilize them to measure task difficulty levels.
Consequently, the optimized meta-learning model can assign gradient weights according to varying task difficulties. With such an enhanced meta-learning process, Meta-Stock can acquire the generalization ability adapted to predict sub-new stock prices more effectively. 
The main contributions of this paper can be summarized as follows:
\begin{itemize}
    \item We propose Meta-Stock, a task-difficulty-adaptive meta-learning approach to address the price prediction problem targeting sub-new stocks, flexible to different backbones. Meta-Stock adapts the generalization ability acquired from old stock price prediction tasks to those of sub-new stocks, thus overcoming sub-new stock data scarcity.
    \item We introduce a task-difficulty-adaptive learning strategy to enhance the meta-learning process. We define task difficulty as price volatility measured by high-frequency coefficients extracted via wavelet transform.%
    \item We show that Meta-Stock outperforms previous methods and demonstrate its applicability in real-world trading via extensive experiments on three stock markets spanning twenty-two years. Given the high profitability of sub-new stocks, Meta-Stock is valuable for stock traders and finance professionals.%
\end{itemize}

\section{Related Work}
\label{sec:2}
\subsection{Stock Price Prediction}
Modern methods based on the Efficient Market Hypothesis \cite{malkiel1989efficient} leverage natural language features to analyze market sentiment \cite{Earningcall2}, supplementing original price data.
The textual features can be extracted from news \cite{NLP2}, social media \cite{Stocknet}, and public earning calls \cite{Earningcall1}.
For instance, Sawhney et al. \cite{NLP1, Earningcall3} propose hierarchical temporal attention and cross-modal attention fusion for NLP-enhanced stock prediction.
The efforts show how natural language data can complement price-based methods in capturing the effect of events like market surprises, mergers and acquisitions over stock returns.

Recent work also attempts to model company relations using stock prices \cite{Graph1,HATS,Graph2} and text data \cite{GraphNLP,MAN-SF} with the GNNs (Graph Neural Networks).
For example, Sawhney et al. \cite{Graph5} and Ang and Lim \cite{ACL22} propose the hyperbolic stock graph attention network and guided attention multimodal multitask network respectively to capture the  inter-company relationship and temporal dependencies in stock prices, promoting accurate stock prediction.

However, despite these competitive results, text-based approaches require a large-scale, high-quality corpus to extract helpful information accurately \cite{dong2020belt,batra2018integrating}. 
The demand in quantity and quality can result in significant time and money.
Moreover, most existing approaches only focus on old stocks that have been listed for over a year and consume substantial training data.
They ignore the significance of sub-new stocks for quantitative trading and thus fail to take the challenge of sub-new stock price prediction into consideration.
To address this problem, we have to uncover the price series' characteristics: volatility, and utilize the valuable information effectively.
The volatility manifests the stock prediction difficulty, which motivates our task-difficulty-adaptive meta-learning design.

\subsection{Meta-learning}
Meta-learning, also known as \textit{learning to learn}, emerges as an efficient method for learning to solve a new task with a limited amount of data by leveraging the generalization capability acquired from previous tasks \cite{hospedales2021meta}.
The idea of meta-learning has been taken to solve the data scarcity problems in many areas, such as recommendation system \cite{Chen2022RecurrentMA} and text classification \cite{Lei2022AdaptiveMV}.
For stock price prediction, Shin{-}Hung et al. \cite{meta-learning} adopt MAML (Model-agnostic Meta-learning) for model training \cite{MAML}. However, despite its great success in solving the data scarcity problem for stock price prediction, MAML has to differentiate through the SGD steps, consuming lots of time. 
Moreover, this method treats tasks with different difficulties equally and fails to consider the inherent volatility of the stock price series \cite{meta-learning}. Unlike the existing method, we choose Reptile, an efficient meta-learning algorithm without differentiating through the SGD steps \cite{Reptile}. To tackle the sub-new stock price prediction problem, we incorporate old stocks and construct meta-learning tasks for the model to acquire a fast generalization ability. We further improve the meta-learning process with an adaptive learning strategy that assigns weights to tasks according to their difficulty measured by volatility.%

\begin{figure*}[!t]
    \centering

\includegraphics[width=\textwidth]{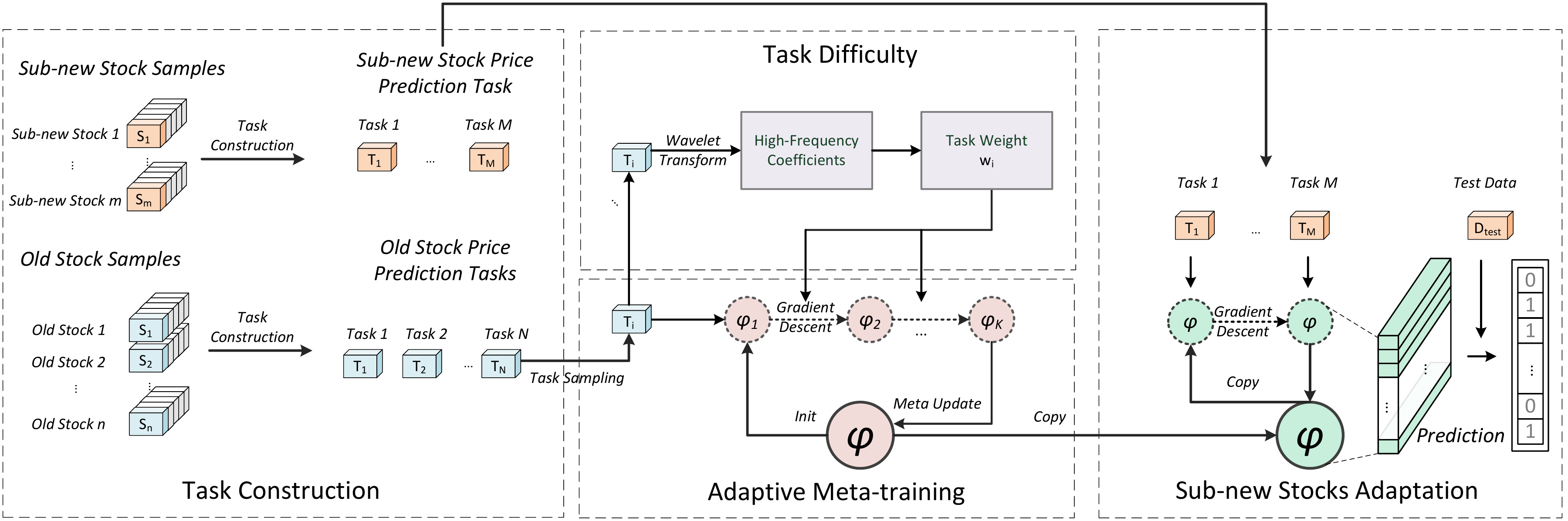}    
    \caption{An overall pipeline of the proposed Meta-Stock, including Task Construction, Task Difficulty, Adaptive Meta-training and Sub-new Stocks Adaptation.}
    \label{fig:model}
\end{figure*}

\section{Methodology}
\label{sec:3}
\textbf{Figure \ref{fig:model}} presents an overview of our proposed Meta-Stock. In the following subsections, we first describe the formulation of the sub-new stock price prediction problem \textbf{(\S\ref{sec:3.1})}. We then articulate the construction of stock price prediction tasks \textbf{(\S\ref{sec:3.2})} and elaborate the quantification of task difficulties \textbf{(\S\ref{sec:3.5})}. Adaptive Meta-training from abundant old stock prediction tasks is introduced in \textbf{(\S\ref{sec:3.3})}, which enables the model to generalize fast across homogeneous tasks. Lastly, we introduce Sub-new Stocks Adaptation to adapt the model with task-agnostic knowledge to sub-new stock prediction with limited samples \textbf{(\S\ref{sec:3.4})}.

\subsection{Problem Formulation}
\label{sec:3.1}
Stock price prediction can be formulated as the time-series classification problem.
Given the $i$-th stock sample $X_i$ in a stock dataset $D$, the stock sample $X_i$ can be denoted as $X_i = \{X_i^1,\dots,X_i^U\}$, where $U$ denotes the time window length of a stock sample. The feature of stock sample  $X_i$ on the $u$-th day can be denoted as $X_i^u \in \mathbf{R}^{d}$, where $u\in(0,U]$ and $d$ denotes the feature dimension of corresponding timestep.
Following \cite{Adv-ALSTM}, the label of stock sample $X_i$ can be defined as: 

\begin{equation}
    Y_i = \{ 
    \begin{matrix}
        0, & p_i^U < p_i^{U-1} \\
        1, & p_i^U \ge p_i^{U-1}
    \end{matrix}
    ,
\end{equation}

where $p_i^u$ denotes the adjusted closing price of $X_i$ on the $u$-th day. $Y_i = 1$ denotes the adjusted closing price rises, and $Y_i = 0$ denotes it drops.
We denote the old stock dataset as $D_\text{old}$ and the sub-new stock dataset as $D_\text{sub-new}$, and our model Meta-Stock aims to acquire the generalization ability from $D_\text{old}$ to facilitate the prediction on $D_\text{sub-new}$.

\subsection{Task Construction}
\label{sec:3.2}
By learning from a diverse set of tasks in the source domain, our model acquires the ability to adapt to target tasks with limited data. %
To achieve this, we devise a task construction strategy that ensures diversity in the meta-training tasks $T_\text{old}={(X_i, Y_i)}_{i=1}^W$ by sampling randomly from old stock data $D_\text{old}$ with different distributions, where $W$ denotes the number of samples in a task.
Likewise, we also construct a few sub-new stock price prediction tasks $T_{\text{sub-new}}={(X_i, Y_i)}_{i=1}^W$.
With the sub-new stock tasks, the meta-learning model can adapt to these tasks after meta-training.
Specifically, the data samples are obtained via sliding window over stock feature series calculated by the adjusted closing price and volume (More details are provided in the \textbf{Section A of the appendix}). 
By updating the model with gradient descent on both old and sub-new stock tasks that share same task sizes $W$, Meta-Stock achieves improved generalization for sub-new stock price prediction.

\subsection{Task Difficulty}
\label{sec:3.5}

The difficulty of stock price prediction tasks lies in the inherent volatility of stock price series.
To measure this, we compute a difficulty score $S_j$ for each training task $T_j \in T$. To determine the scores $S_j$, we calculate a sample difficulty score $S_i$ for each data sample $X_i \in T_j$ with the following approach.
For each data sample $X_i \in T_j$, $S_i$ reflects the temporal volatility in price and volume signals present in $X_i$. However, quantifying price volatility in the time domain is challenging. Therefore, we use wavelet transform techniques to analyze the volatility in the frequency domain instead.

One common approach is the Fourier transform (FT) \cite{FT}, which creates a representation of the signal in the frequency domain. However, the Wavelet transform (WT) \cite{WT} provides more localized information of the signal in both the time and frequency domains. Hence, we employ the Discrete Wavelet Transform (DWT) \cite{DWT} to decompose the multivariate time series stock sample $X_i$ into its smooth (low-frequency) coefficients $L_{\lambda,\mu}$ and its detail (high-frequency) coefficients $H_{\lambda,\mu}$.

During DWT, the original multivariate time series $X_i$ is convolved with a low-pass filter and a high-pass filter, and their outputs are downsampled to obtain the smooth (low-frequency) coefficients $L_{\lambda,\mu}$ and the detail (high-frequency) coefficients $H_{\lambda,\mu}$, respectively. The frequency-domain volatility of the time series $X_i$ can now be quantified with the DWT coefficients as follows: 
\begin{equation}
    L_{\lambda,\mu}=\int X_i \Phi_{\lambda,\mu}(t) d t
\end{equation}
\begin{equation}
    H_{\lambda,\mu}=\int X_i \Psi_{\lambda,\mu}(t) d t
\end{equation}
where $\Phi$ and $\Psi$ are, respectively, the father and mother wavelets, and $\lambda$ and $\mu$ are, respectively, the scaling and translation parameters. The father wavelet approximates the smooth (low-frequency) components of the signal, and the mother wavelet approximates the detail (high-frequency) components. The father wavelet $\Phi$ and the mother wavelet $\Psi$ are defined as follows:
\begin{equation}
    \Phi_{\lambda,\mu}(t)=2^{-\lambda/2} \Phi(2^{-\lambda}t-\mu) d t
\end{equation}
\begin{equation}
    \Psi_{\lambda,\mu}(t)=2^{-\lambda/2} \Psi(2^{-\lambda}t-\mu) d t
\end{equation}
The two wavelets $\Phi$ and $\Psi$ satisfy the following condition:
\begin{equation}
    \int \Phi(t) d t = 1
\end{equation}
\begin{equation}
    \int \Psi(t) d t = 0
\end{equation}
The detail coefficients along the temporal dimension contain high-frequency information and indicates the volatility and associated risk of the financial time-series \cite{high-freq}.
We thus quantify the sample difficulty $S_i$ with volatility measured by high-frequency components after discrete wavelet transform $c_{\lambda,\mu}^{i}$.
Once we obtain every sample's difficulty $c_{\lambda,\mu}^i$ in the task $T_j$, we can measure the task difficulty $S_j$ by their root sum of squares:
\begin{equation}
    S_j = \sqrt{ \sum_{i}^{W}{c_{\lambda,\mu}^i}^2}
\end{equation}
where the task $T_j$ contains $W$ samples.

\subsection{Adaptive Meta-training}
\label{sec:3.3}
As \textbf{Figure \ref{fig:model}} shows, our model learns from numerous old stock price prediction tasks to extract task-agnostic knowledge and acquires the fast generalization capability, which can be measured by the model's average predicting performance for $K$ meta-training steps on task $T_j$.
Therefore, we establish the objective of Adaptive Meta-training to minimize the expected loss given a selected task $T_j$:

\begin{equation}
    \min_{\varphi}{\mathbb{E}_{T_j}}[L_{\varphi}(U_{T_j}^{K}(\varphi))]
\end{equation}

where $U_{T_j}^{K}(\varphi) = \varphi_k$, the model learning on task $T_j$. 
When learning on task $T_j$, Reptile optimizes the model $\varphi_k$, where $k\in(0, K]$, with gradient descent for $K$ meta-training steps as follows: 
\begin{equation}
    \varphi_k = \varphi_k - \alpha \bigtriangledown_{\varphi_k} L_{T_j} (f_{\varphi_k})
\end{equation}
where $\alpha$ denotes the fixed learning rate and $L_{T_j}$ represents the loss on task $T_j$. 
In contrast to the Reptile algorithm, Meta-Stock aims to capture a better learning strategy of stock price prediction by assigning weights $w_j$ to a given stock price prediction task $T_j$ according to varying task difficulty $S_j$.
\begin{equation}
\label{eq: Loss}
    \varphi_k = \varphi_k - \alpha w_j \bigtriangledown_{\varphi_k} L_{T_j} (f_{\varphi_k})
\end{equation}
However, if we retain the extreme values or outliers in weights $w_j$, the weighted gradients can be too large, and thus bypass the local minimum and overshoot. Otherwise, it can be too small, and hence increases the total computation time to a very large extent.
Therefore, we normalize the task difficulty $S = [S_1,\dots, S_N]$ by a softmax function to get a weight vector $w = [w_1,\dots, w_N]$ for all the old-stock tasks $T_\text{old} = [T_1,\dots, T_N]$, where there are $N$ old-stock tasks in total.
The softmax normalization is a way of reducing the influence of extreme values or outliers in the weight vector without removing data points from the set. For task $T_j$, the weight $w_j$ of this task is computed with all $W$ samples in $T_j$ as:
\begin{equation}
    w_j = \frac{e^{S_j}}{\sum_i^W e^{S_i}}
\end{equation}

In Equation \ref{eq: Loss}, we train the model by minimizing the cross-entropy loss $L_{T_j}$, given as:
\begin{equation}
    L_{T_j} = -\sum_{i=1}^{W} Y_{i} \ln \left(y_{i}\right)+\left(1-Y_{i}\right) \ln \left(1-y_{i}\right)
\end{equation}
where $Y_i$ denotes the true price movement of a stock sample $X_i$ from  the training data of the tasks $T_j$. $y_i$ denotes the prediction of the model $\varphi_k$ for the stock sample $X_i$.
After learning from the tasks $T_j$, we can optimize the meta-learning objective as shown below:
\begin{equation}
    \varphi = \varphi + \beta(\varphi_k - \varphi)
\end{equation}
where $\beta$ denotes the meta-learning rate.
Here, we aggregate $K$ meta-training task gradients to obtain a meta-gradient $\varphi_k - \varphi$. With the meta-gradient, we move the initial parameter of the model $\varphi$ in the direction of the average of the task model parameters $\varphi_k$. 
Hence, the model converges towards a solution $\varphi_k$ close (in Euclidean distance) to each task $T_j$ ’s manifold of optimal solutions \cite{Reptile}. 
Because the meta-learning model parameters $\varphi$ are close to the optimal parameters of each task $T_j$, only a few gradient updates are required to obtain the optimal solutions for each task $T_j$. 
Therefore, Meta-Stock enables the model $\varphi$ to generalize on different tasks with task-agnostic knowledge.
To show more details, we outline the optimization process in \textbf{Algorithm \ref{algo:Meta-Stock}}. 

\SetAlgoNoLine
\begin{algorithm}[t]
\SetKwInOut{Require}{Require}
\caption{Adaptive Meta-training} \label{algorithm}
\label{algo:Meta-Stock}
\SetNlSty{}{}{:}
    \Require{$Z(T)$: distribution of the task T}
    \Require{$\alpha$, $\beta$: learning rate hyperparameters}
    \Require{$w_j$: weight measured by task $T_j$'s difficulty}
        randomly initialize $\varphi$, the vector of initial parameters \\
                \For {\text{\textbf{\upshape all}} $T_j \sim Z(T)$}{
                    $\varphi_1 \gets \varphi$ \\
                    \For {\text{\upshape every meta training step} $k$}{
                        Evaluate $\bigtriangledown_{\varphi_k} L_{T_j} (f_{\varphi_k})$ with respect to task samples \\
                        Compute adapted parameters with gradient descent: $\varphi_k = \varphi_k - \alpha w_j \bigtriangledown_{\varphi_k} L_{T_j} (f_{\varphi_k})$ \\
                    }
                    Update $\varphi = \varphi + \beta(\varphi_k - \varphi_1)$
                }
\end{algorithm}

\subsection{Sub-new Stocks Adaptation}
\label{sec:3.4}
After acquiring the generalization ability on old-stock tasks, our model $\varphi$ can generalize efficiently to sub-new stock tasks with a handful of training data via a few gradient steps and obtain the adapted parameters. 
This fast adaptation comes from the fact that we have already simulated fast learning on multiple tasks with limited data in the Adaptive Meta-training phase.
In particular, we minimize the loss in $\varphi$ on the sub-new stock price prediction tasks through gradient descent. 

\begin{equation}
    \varphi=\varphi-\gamma \nabla_\varphi L_{T_\text{sub-new}}(\varphi)
\end{equation}
where $L_{T_\text{sub-new}}$ denotes the cross-entropy loss on the sub-new stock price prediction task and $\gamma$ refers to the learning rate.

\section{Experiments and Setup}

\begin{table*}[!htp]\centering
\caption{Meta-Stock performance comparison against baseline models and methods. Except Train on sub-new stocks, all methods introduce old stock data to assist sub-new stock prediction. Values in \textbf{bold}, \underline{underline} denote \textbf{best} and \underline{second-best} results, respectively. $*$ indicates improvements over the same backbone but with other methods are statistically significant ($p < 0.01$), under Wilcoxon’s signed rank test.}\label{table:bigtable}
\scriptsize
\resizebox{\textwidth}{!}{
\begin{tabular}{llcccccccccc}\toprule
\multirow{2}{*}{Method} &\multirow{2}{*}{Backbone} & &\textbf{US-STOCKS} & & &\textbf{CN-STOCKS} & & &\textbf{HK-STOCKS} & \\\cmidrule{3-5}\cmidrule{6-8}\cmidrule{9-11}
& &ACC &MCC &F1 &ACC &MCC &F1 &ACC &MCC &F1 \\\midrule
\multirow{4}{*}{Train on sub-new stocks} &LSTM-FCN &56.29 &4.00 &52.66 &51.50 &3.69 &46.21 &50.54 &1.34 &43.92 \\
&ResCNN &53.24 &-0.70 &51.25 &51.75 &0.89 &46.71 &51.24 &-2.99 &46.31 \\
&ResNet &55.19 &3.38 &53.08 &53.72 &4.90 &51.48 &54.93 &0.57 &47.78 \\
&InceptionTime &53.67 &-0.42 &51.25 &51.30 &0.46 &46.58 &51.18 &-0.23 &45.44 \\
\cmidrule{1-11}
\multirow{4}{*}{Transfer Learning} &LSTM-FCN &\underline{56.31} &4.09 &52.73 &53.91 &6.69 &53.61 &53.65 &0.23 &49.95 \\
&ResCNN &53.71 &3.77 &53.41 &54.03 &7.56 &54.02 &53.34 &-0.30 &49.83 \\
&ResNet &56.23 &4.80 &53.43 &54.61 &7.42 &53.05 &55.61 &1.66 &47.18 \\
&InceptionTime &55.49 &3.05 &52.64 &55.07 &8.25 &53.39 &53.77 &1.75 &51.28 \\
\cmidrule{1-11}
\multirow{4}{*}{Reptile} &LSTM-FCN &55.65 &6.54 &54.83 &56.17 &10.27 &53.98 &54.66 &3.94 &52.38 \\
&ResCNN &55.71 &2.54 &52.01 &55.94 &10.09 &54.72 &51.79 &1.44 &51.65 \\
&ResNet &55.00 &4.99 &54.09 &55.53 &8.91 &53.56 &53.01 &4.22 &53.56 \\
&InceptionTime &\textbf{56.85} &5.04 &52.80 &55.17 &8.40 &53.80 &\underline{56.61} &3.48 &41.10 \\
\cmidrule{1-11}
\multirow{4}{*}{Meta-Stock} &LSTM-FCN &55.90 &\underline{9.42}$^{*}$ &\textbf{55.93}$^{*}$ &56.23$^{*}$ &10.35$^{*}$ &53.86 &53.92 &6.39$^{*}$ &53.95$^{*}$ \\
&ResCNN &54.89 &8.59$^{*}$ &55.13$^{*}$ &56.39$^{*}$ &10.90$^{*}$ &\underline{54.77} &56.47$^{*}$ &\underline{8.48}$^{*}$ &\textbf{54.75}$^{*}$ \\
&ResNet &54.78 &\textbf{9.68}$^{*}$ &55.09$^{*}$ &\textbf{56.75}$^{*}$ &\textbf{11.47}$^{*}$ &53.71 &\textbf{57.28}$^{*}$ &\textbf{9.49}$^{*}$ &\underline{54.65}$^{*}$ \\
&InceptionTime &56.26 &8.86$^{*}$ &\underline{55.90}$^{*}$ &\underline{56.48}$^{*}$ &\underline{11.33}$^{*}$ &\textbf{55.47}$^{*}$ &56.35$^{*}$ &6.71$^{*}$ &52.83$^{*}$ \\
\bottomrule
\end{tabular}
}
\end{table*}

\subsection{Dataset}
\label{sec:4}
For the dataset, we choose the stock markets in US, mainland China and Hong Kong due to their large capitalization and numerous companies. 
We then collect the dataset from AKShare \cite{akshare2019} on the \textit{three} real-world stock markets, from 01/01/2000 to 22/02/2022 and denote the market as US-STOCKS, CN-STOCKS and HK-STOCKS, respectively.
We preprocess data and shift a 5-day lookback window along the trading days to generate samples following \cite{NLP1}. We label and filter the samples based on the movement percentage of the adjusted closing price. For example, samples with movements $\geq$ 0.55$\%$ and $\leq$ -0.5$\%$ are labeled as positive and negative, respectively, and the remained samples are filtered out. To distinguish old and sub-new stocks, we split the dataset into old stock datasets ranging from 01/01/2000 to 22/02/2021 and sub-new stock datasets ranging from 23/02/2021 to 22/02/2022. We further split the sub-new stock datasets in the proportion of 6:2:2 and obtain training data from 22/02/2021 to 22/09/2021, validation data from 23/09/2021 to 02/12/2021 and test data from 03/12/2021 to 22/02/2022. 
The detailed statistics of our proposed dataset are shown in the \textbf{Section A of the appendix}.

\subsection{Training Setup}
We perform all experiments on an Nvidia GeForce GTX 1080Ti GPU.
We train Meta-Stock for 50 epochs with AdamW optimizer. 
We use grid search to find optimal hyperparameters for Meta-Stock based on validation performance.
We set the length of stock sample $T=5$, training steps $K=6$, meta batch size $B=6$, batch size $C=4096$, weight decay rate $\sigma=1e-5$, learning rate $\alpha, \beta, \gamma \in (1e-4, 1e-1)$ for Meta-Stock. Here, the number of samples in a task $W=24576$, which is equal to $B*C$ and the number of old stock tasks $N$ and sub-new stock tasks $M$ can be calculated with $W$, the total number of old stock samples and the total number of sub-new stock samples.
We repeat each experiment 5 times and record the average performance. 
For model evaluation in stock price prediction, we follow the metrics in \cite{Adv-ALSTM} calculate the Accuracy (ACC), Matthews Correlation Coefficient (MCC) and F1-score (F1).
For model evaluation in stock trading, we choose the Annual Return Ratio (ARR), Sharpe Ratio (SR) \cite{Sharpe1994TheSR}, Maximum Drawdown (MDD) \cite{magdon2004maximum}, Sortino Ratio (SoR) \cite{sortino1994performance}, Calmar Ratio (CR) \cite{young1991calmar} and Omega Ratio (OR) \cite{keating2002universal}.
We formulate the evaluation metrics and explain their details in the \textbf{Section A of the appendix}.

\subsection{Baselines and Backbones}
We choose the following baseline approaches to train different backbones and compare their performances with Meta-Stock:
\begin{itemize}[label={\LARGE\textbullet}]
    \item \textbf{Train on sub-new stocks}: The backbones are trained on the training set of sub-new stocks and then tested on the test set of sub-new stocks.
    \item \textbf{Transfer Learning}: The backbones are pre-trained on the old stocks and then finetuned on the training set of sub-new stocks. Finally, we test the backbones on the test set of sub-new stocks.
    \item \textbf{Reptile}: The backbones are meta-trained on the old stocks and then adapted to the training set of sub-new stocks. Finally, we test the backbones on the test set of sub-new stocks. %
\end{itemize}
For backbones, we choose LSTM-FCN \cite{vceponis2020investigation}, ResCNN \cite{ResCNN}, ResNet \cite{li2020enhancing} and InceptionTime \cite{wang2021hierarchical}. We provide backbone details in the \textbf{Section A of the appendix}. Note that recent NLP-based stock price prediction models are not considered as comparison due to the limited availability of text data for sub-new stocks.

\section{Results and Discussion}
\subsection{Performance Comparison}
\label{sec:4.1}
We compare Meta-Stock with various approaches.
Meta-Stock achieves a state-of-the-art performance in terms of ACC, MCC, and F1 as shown in \textbf{Table \ref{table:bigtable}}. 
Moreover, following \cite{GraphNLP}, we employ Wilcoxon’s signed rank test \cite{conover1999practical} and reveal significant improvements $(p<0.01)$ of Meta-Stock over the compared methods.
With such an advance, Meta-Stock validates its effectiveness, though facing both bullish and bearish conditions in three markets.
We attribute the improvement of Meta-Stock over other approaches to three reasons.
First, Meta-Stock formulates the sub-new stock prediction problem from a new task-based perspective, allowing our model to learn from various tasks and capture a better learning strategy on stock price prediction tasks. With the mastered strategy, Meta-Stock improves generalization across homogeneous tasks and thus learns faster than many state-of-the-art methods. 
Second, we design a strategy to construct training tasks with various data distribution, which enable Meta-Stock to better learn the homogeneous data pattern in different distributions.
By perceiving similar data patterns between old and sub-new stocks, Meta-Stock can utilize old stock data more efficiently and generalize better to sub-new stock price prediction.
Third, when finetuning the meta-learning model on the sub-new stock data, we keep the training strategy on sub-new stock data the same as that on old stock data, which enable the model to better apply the obtained task-agnostic knowledge to the prediction of sub-new stocks.

\subsection{Profit Analysis}
\label{sec:4.4}
\begin{table}[!htp]\centering
\caption{Real-world trading analysis using Meta-Stock across stocks in US, CN, HK stock markets. We observe significantly high profit improvements and risk reduction when we use Meta-Stock to train ResNet for trading.}
\scriptsize
\resizebox{\columnwidth}{!}{
\begin{tabular}{lccccccc}\toprule
Method &AR &MDD &SR &SoR &CR &OR \\\midrule
Reptile &-0.793 &-0.801 &-2.084 &-2.165 &-0.99 &0.449 \\
\textbf{Meta-Stock} &\textbf{1.413} &\textbf{-0.205} &\textbf{2.048} &\textbf{6.653} &\textbf{6.744} &\textbf{1.805} \\
\bottomrule
\end{tabular}
}
\label{tab: profit}
\end{table}

We examine the practical applicability of Meta-Stock to real-world stock trading by analyzing the pure returns (Annual Return Rate), risk-adjusted returns (Sharpe ratio, Sortino Ratio, Calmar Ratio, Omega Ratio) and the maximum risk (Maximum Drawdown) associated with the trades using ResNet across stocks in US, CN and HK markets. We follow a trading strategy: if the model predicts the rise of a stock’s price the next day, we will buy the stock at the closing price and sell it at the closing price when the model speculates a price fall. We first train ResNet with Reptile, which is Meta-Stock without adaptive learning for stock trading, and observe poor performance in terms of profits and a high risk for all markets as shown in \textbf{Table \ref{tab: profit}}.  This observation indicates that Reptile takes riskier trading decisions and often experiences enormous losses. However, when we train ResNet using Meta-Stock, we observe significant improvements in risk-adjusted returns (781.21\%) and a substantial reduction in maximum losses (73.75\%). Such improvements indicate the efficacy of Meta-Stock in enhancing the real-world applicability of neural stock prediction methods. We further elucidate the benefits of Meta-Stock via a qualitative study.

\subsection{Probing Task Difficulty}
\begin{table}[!htp]\centering
\caption{Performance improvements after training for 1 epoch across task groups with varying difficulty scores ($S$): validating the task difficulty measurement.}\label{tab: difficulty}
\scriptsize
\resizebox{\columnwidth}{!}{%
\begin{tabular}{ccc}\toprule
Task Difficulty Score ($S$) & MCC Relative Gains in CN-STOCKS (\%) \\\midrule
Easy &1.06\% \\
Medium &0.88\% \\
Hard &-0.53\% \\
\bottomrule
\end{tabular}
}
\end{table}
In this study, we investigate the performance improvements achieved by training on samples of varying difficulty levels for 1 epoch in the beginning.  To this end, we divide our dataset into three groups of tasks with different levels of difficulty: easy, medium, and hard. Specifically, we distribute three task groups with the same amount of tasks by their difficulty scores. For instance, the tasks with the top 1/3 difficulty scores are assigned to hard groups, and those with the bottom 1/3 are assigned to easy groups. The resulting performance gains obtained through the Reptile algorithm are presented in \textbf{Table~\ref{tab: difficulty}}.
Our results highlight the effectiveness of Reptile in improving the performance of stock price prediction tasks over data with varying levels of difficulty, with improvements observed for both easy and medium difficulty levels.  However, the algorithm exhibits a decline in predicting ability for the hard-level group of tasks.
Interestingly, we also note that at the beginning of the learning process, the relative improvement for stock price prediction increases as the task difficulty decreases from hard to medium to easy.  These findings are consistent with the typical learning curve of humans, that is, learning with increasing difficulty. 
Learning complex tasks ahead of time can be frustrating for humans if they cannot solve simple tasks.
Therefore, the observations validate the effective quantification of task difficulty.

\subsection{Analyzing the Effectiveness of Meta-Stock}
\begin{table}[!htp]\centering
\caption{ACC improvements of Meta-Stock+ResNet over Reptile+ResNet across groups of tasks having varying task difficulty scores ($D$).}\label{tab: diff1}
\scriptsize
\resizebox{\columnwidth}{!}{%
\begin{tabular}{ccccc}\toprule
\multirow{2}{*}{Task Difficulty ($D$)} &\multicolumn{3}{c}{ACC Relative Gains (\%)} \\\cmidrule{2-4}
&US-STOCKS &CN-STOCKS &HK-STOCKS \\\midrule
Easy &3.36\% &7.64\% &2.69\% \\
Medium &2.27\% &6.07\% &9.44\% \\
Hard &3.93\% &5.70\% &8.15\% \\
\bottomrule
\end{tabular}
}
\end{table}
We now study the performance improvements obtained via Meta-Stock over Reptile against samples of varying difficulty levels.  
In \textbf{Table~\ref{tab: diff1}} we divide the dataset into groups of easy, medium, and hard tasks according to the task difficulty $D$.  We observe significant improvements over all three difficulty levels on all evaluation metrics, demonstrating that Meta-Stock improves performance across sub-new stock price prediction tasks with varying difficulty levels (more results about improvements on the MCC and F1 scores can be referred to in the \textbf{Section B of the appendix}).  We contribute these improvements to Meta-Stock's adaptive learning strategy that assigns more weight to complicated tasks. Once Meta-Stock can better handle complicated tasks, the easier ones can also be solved better. 

\begin{figure}[!t]
    \centering
    \includegraphics
        [width=\columnwidth]
        {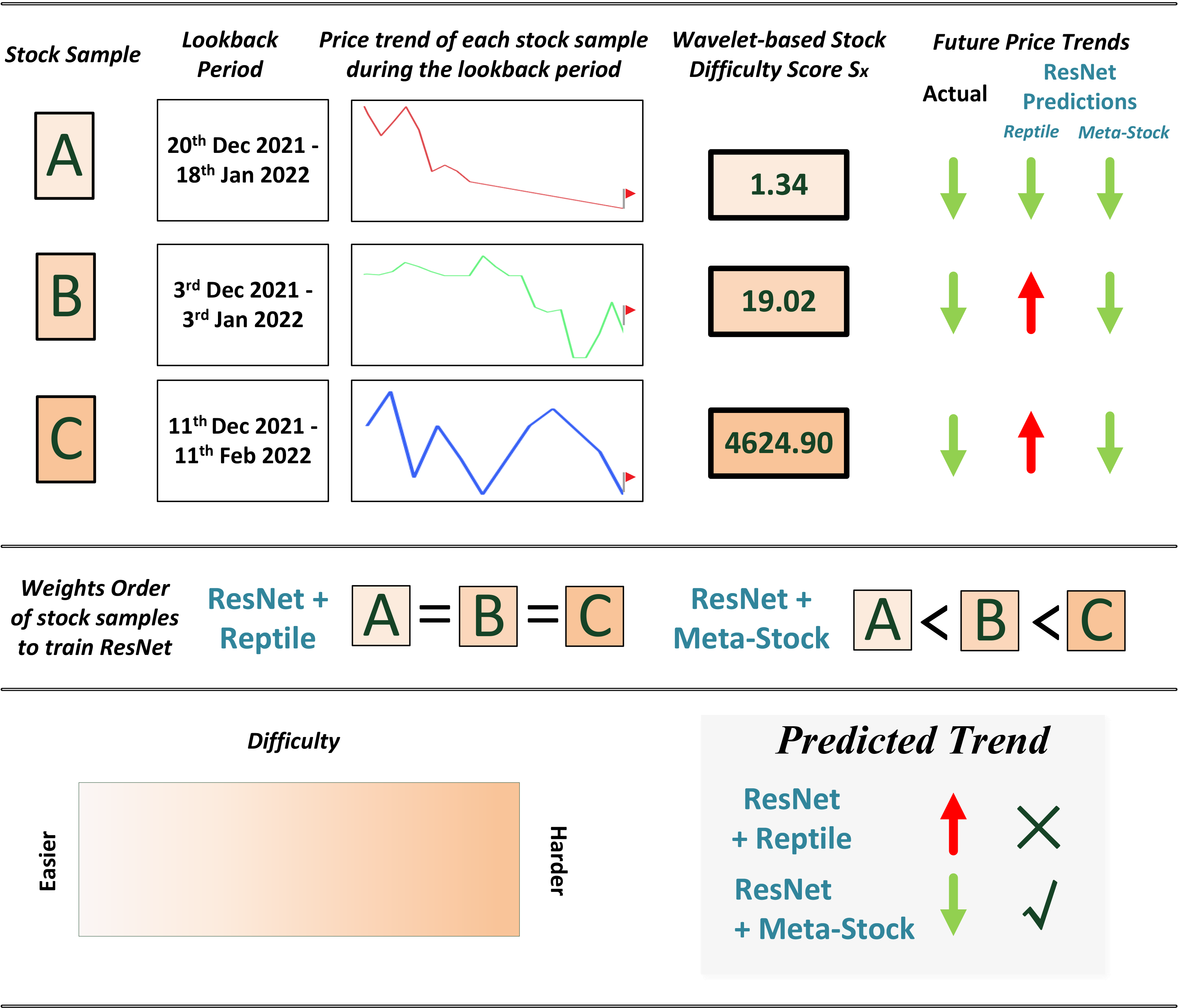}
    \caption{Case study on the US, China, HongKong markets showing how Meta-Stock allows ResNet to learn across stock examples with adaptive weights according to their difficulties. This enhances model training and stock prediction performance.
    }
    \label{tab:risk}
\end{figure}

\subsection{Qualitative Analysis}
\label{sec:4.5}
We further conduct an extended study to elucidate the benefits of Meta-Stock for stock prediction, as shown in Figure \ref{tab:risk}.
Price series in training sample C possess a volatile trend, making it hard to analyze the future trend of the stock. We show that for a moderately complex test-data sample C, its movement is incorrectly classified when training ResNet without the adaptive learning. However, when training with Meta-Stock, its price trend is classified accurately. We attribute Meta-Stock+ResNet’s overall improved performance to the generated task weights that ameliorates the efficiency of the learning process.

\section{Conclusion}
In this paper, we propose Meta-Stock, a task-difficulty-adaptive meta-learning approach to predict sub-new stock price trends. 
Our meta-learning approach seeks to solve the data scarcity of sub-new stocks by leveraging old stocks and acquiring the fast generalization ability that can be extended to sub-new stock price prediction. Furthermore, we improve the entire meta-learning process by introducing adaptive learning according to volatility levels.
We display Meta-Stock's applicability in sub-new stock price prediction and real-world trading through extensive quantitative and qualitative experiments on real market data. 
In future work, we intend to extend Meta-Stock’s architecture to enhance its scalability in cross-market scenarios.

\newpage
\bibliography{aaai24}

\end{document}